\documentclass[12pt]{article}
\usepackage{epsfig,amssymb}

\def\ltsim{\lower3pt\hbox{$\, \buildrel < \over \sim \, $}}
\def\gtsim{\lower3pt\hbox{$\, \buildrel > \over \sim \, $}}
                                                                                                                                                                                                                                                                                                                                                                                                                                                                                                                                                                                                            \def\be{\begin{equation}}
\def\ee{\end{equation}}
\def\ba{\begin{eqnarray}}
\def\ea{\end{eqnarray}}
\def\ga{\mathrel{\raise.3ex\hbox{$>$\kern-.75em\lower1ex\hbox{$\sim$}}}}
\def\la{\mathrel{\raise.3ex\hbox{$<$\kern-.75em\lower1ex\hbox{$\sim$}}}}

\begin{document}

\baselineskip=16pt
\begin{titlepage}

\begin{center}

\vspace{0.5cm}

\Large {\bf On Stability Of The Crystal Universe Models
}\\

\vspace{10mm}
Yun-Song Piao$^{a}$, Xinmin Zhang$^{b}$ and Yuan-Zhong Zhang$^{c,a}$ \\

\vspace{6mm}
{\footnotesize{\it
$^a$Institute of Theoretical Physics, Chinese Academy of Sciences, \\
 P.O. Box 2735, Beijing 100080, China \footnote{Mailing address in China}\\
$^b$Institute of High Energy Physics, Chinese Academy of Sciences, P.O.
Box 918(4), Beijing 100039, China
\\
$^c$CCAST (World Lab.), P.O. Box 8730, Beijing 100080}\\}

\vspace*{5mm}
\normalsize


\smallskip
\medskip


\smallskip
\end{center}
\vskip0.6in

\centerline{\large\bf Abstract}

We generalize the Goldberger-Wise mechanism and study the stability of the
Crystal Universe models. We show that the model can be
stabilized, however
for configurations of Crystal Universe in the absence of fine-tuning,
brane crystals are not equidistant, {\it
i.e.} a $"-+"$ pair is far away
from adjacent $"-+"$ pair, except for the fixed points of the orbifold,
which differs from
 the assumptions taken
in the literature.

\vspace*{2mm}

\end{titlepage}


Recently there has been considerable interest in studying the brane
universe models. Randall and Sundrum (RS) \cite{RS} have proposed a high
dimensional picture to
solve the gauge hierarchy problem between the Planck scale and the
electroweak scale. In RS's scenario there are two 3-branes
with opposite
tension which sit at the fixed points of an $S^1/Z_2$ orbifold with
$AdS_5$ bulk geometry,
the gravity is shown to be localized on the brane
of positive
tension ($"+"$ brane)
and the exponential warp factor in the spacetime metric generates a
scale hierarchy on the brane of negative tension ($"-"$ brane). If the
standard model (SM) fields
reside on the $"-"$ brane, this leads to the resolution of the
gauge hierarchy problem, however
even though it is plausible in string
theory, it is still questionable whether we can live on
$"-"$ brane.
There are many variants of RS model proposed recently to avoid this
problem. In Ref.\cite{LR},
Lykken and Randall \cite{LR}
proposed a multi-brane model, i.e. a $"++-"$ brane
configuration in which the SM fields reside on intermediate $"+"$ brane
and a warp factor
accounts for scale hierarchy. Kogan et al. in Ref.\cite{KM1} (see also
\cite{KM2} and \cite{GR} ) considered $"+--+"$
multi-brane model and the Crystal Universe model
(see also \cite{O}, \cite{HS}, \cite{l}, \cite{K}, \cite{li}).
These models provide
a way of having the visible sector on a $"+"$ brane with an hierarchical
warp factor and interestly predict gravity be different from
we expect not only at small scale but also at ultralarge scale.
Furthermore it has
been argued in \cite{ADDK} that this class of models give rise to many new
phenomena such as neutrino mixing, Dark Matter  which can be tested
experimentally.

In this paper we study the issue of stability of the crystal
universe models.
Following
Goldberger and Wise (GW) \cite{GW} we introduce
a bulk scalar field into the models, then minimize the potential
generated
by the bulk scalar with quartic interaction localized on two 3-branes.
We will show that the brane crystals is not equidistant in the absence
of fine-tuning and
generally a $"-+"$
pair is far away
from adjacent $"-+"$ pair.


To begin with, we consider a
Crystal Universe Model shown in Fig.1 which
consists of $n$ array of parallel 3-branes with $"+"$ brane every other
$"-"$ brane in a $AdS_5$ space
with negative cosmological constant $\Lambda$. The fifth dimension $y$ has
orbifold
geometry $S^1/Z_2$.
The $n+1$ array of parallel 3-branes are
located at $y_0=0$, $y_1$, $y_2$ ... $y_n$,
where $y_0=0$ and $y_n$ are orbifold fixed points. The action for this
configuration is
\be
S = \int d^4x\int dy
\sqrt{G}\left(2M^3R-\Lambda\right)-\sum_{i-1}\int_{y=y_{i-1}}d^4x V_{i-1} \sqrt{
g^{(i-1)}},
\ee
where $i-1=0, 1, 2, ... n$,
$g^{(i-1)}_{\mu\nu}$ are the induced metric on the branes, $V_{i-1}$
are their tensions and $M$ is the 5D fundamental scale.
The 5D metric ansatz that respects 4D Poincare invariance is given by
\be
ds^2 = e^{-2\sigma(y)}\eta_{\mu\nu}dx^{\mu}dx^{\nu} - dy^2,
\ee
here the warp function $\sigma(y)$ is essentially a conformal
factor that rescales the 4D component of the metric.
Substituting $(2)$ into the Einstein equations
we have that:
\ba
\left(\sigma '\right)^2&=&k^2, ~~~~~~~
\sigma ''= \sum_{i-1}\frac{V_{i-1}}{12M^3}\delta(y-y_{i-1}),
\ea
where $k=\sqrt{\frac{-\Lambda}{24M^3}}$ is effectively the bulk curvature
in the region between the two adjacent brane crystals.
There are two solution to equations $(3)$:
\be
\textrm{ (i).\,\,\,\,}
\sigma_i(y)=(-1)^{i+1} ky+\sum_{j=0}^{i-1} 2 (-1)^{j+1} ky_{j},
\ee
and
\be
\textrm{(ii).\,\,\,\,\,\,\,\,\,\,\,\,\,}
\sigma_i(y)=(-1)^{i} ky+\sum_{j=0}^{i-1} 2 (-1)^{j} ky_{j},
\ee
here $\sigma_i(y)$ are warp factors between the $(i-1)$th brane and $i$th brane.
For solution (i) the $"+"$ brane sits on the fixed point $y_0=0$ and
the corresponding brane tensions are
$V_{i-1}=(-1)^{i-1}\Lambda/k$;
for
solution (ii) on the fixed point is
the $"-"$ brane and
 $V_{i-1}=(-1)^{i}\Lambda/k$.

To study the stability of Crystal Universe, we introduce and
couple a bulk scalar
field to the brane crystals. This technique is a
generalization of the GW
mechanism, however the calculation invlolved in this paper will be much more
complicated than that in \cite{GW}. For a bulk scalar field with mass $m$,
\be
S_{\rm Bulk}= {1\over 2}\int d^4x \int dy
\sqrt{G}\left(G^{AB}\partial_A\Phi\partial_B\Phi -m^2\Phi^2\right),
\ee
where $G_{AB}$ is the 5D metric given in $(2)$ with
$\sigma(y)$ given in $(4)$ and $(5)$.
And the boundary potentials of scalar field are
\be
S_{i-1}=-\int d^4x \lambda_{i-1} \sqrt{
g^{(i-1)}}\left(\Phi^2-v_{i-1}^2\right)^2 ,
\ee
where $v_{i-1}$ are the vacuum expectation values of bulk scalar field
in the $(i-1)$th brane,
$\lambda_{i-1}$ are coupling contants.
We are interested in those configurations of the bulk scalar where
the boundary
potentials are minimised. This essentially amounts
to negligible dynamics of $\Phi$ along the direction tangential to any of the
3-branes. This assumption is reasonable because we focus on the
stability of the 3-branes system at the moment and do not study
phenomenologcal consequence of possible coupling of the bulk scalar field
$\Phi$ to matter fields living on the branes. It, therefore, suffices to
concentrate on equation of motion of $\Phi$ only in $y$ direction,
which is
\be
\partial^2_y\Phi -4\sigma'_i(y)\partial_y\Phi - m^2\Phi =0 ,
\ee
where $\sigma'_i(y) = d\sigma_i(y)/dy$.
Solution of this eqution is
\be
\Phi(y) = \exp{(2\sigma'_i(y)y)}[A_i \exp{(\sigma'_i(y)\nu_i
y)}+B_i\exp{(-\sigma'_i(y)\nu_i y)}] .
\ee
In Eq.(9) $\nu_i = \sqrt{4+m^2/\sigma^{\prime 2}(y)}$ is
independent of $y$ (which we denote by $ \nu$ in the following
discussion),
however
$A_i$ and $B_i$ vary in the range of  $y_{i-1} < y < y_i$.

To determine the coefficients $A_i$ and $B_i$, we require
 $\Phi$
minimize the boundary potential.
We firstly consider the case (i) where
the $"+"$ brane is at $y_0=0$, then discuss the case (ii).

For case (i):
\be
A_i = ({v_{i-1} -v_i R_{i}^{2-\nu}\over 1-R_{i}^{-2\nu}}) Y_{i-1}^{2+\nu},
\ee
\be
B_i = ({-v_{i-1} R_i^{-2\nu}+v_i R_{i}^{2-\nu}\over 1-R_{i}^{-2\nu}})
Y_{i-1}^{2-\nu},
\ee

where $i=2j-1$ with $j=1,2,3 ... $

and
\be
A_i = ({v_{i-1} -v_i R_{i}^{\nu -2}\over 1-R_{i}^{2\nu}}) Y_{i-1}^{-(2+\nu)},
\ee
\be
B_i = ({-v_{i-1} R_i^{2\nu}+v_i R_{i}^{\nu -2}\over 1-R_{i}^{2\nu}}) Y_{i-1}^{\nu -2},
\ee

where $i=2j$ with $j=1,2,3 ... $

In Eqs.(10-13), $Y_i$ and $R_i$ are defined as
:
$Y_i=\exp{(-k y_i)}$,
$R_1=\exp{(-k y_1)}\equiv Y_1$, and
$R_i=\exp{[-k(y_i-y_{i-1})]}\equiv Y_i/Y_{i-1}$ $(i\neq 1)$.

Substituting $A_i$, $B_i$ in Eqs.(10-13) and $\Phi(y)$ in Eq.(9)
into the action $(6)$ and integrating out $y$ give rise to a
4D effective potential $V(R_i, v_i)$,
\begin{eqnarray}
&&k^{-1}V(R_i, v_i)=f_1(R_1, v_0, v_1)+{R_1^4 \over R_2^4}f_2(R_2, v_2, v_1)
+{R_1^4 \over R_2^4}[f_3(R_3, v_2, v_3)
\nonumber \\
&+&{R_3^4 \over R_4^4}f_4(R_4, v_4, v_3)]
+ ... +{R_1^4 R_3^4 ... R_{2j-3}^4\over R_2^4 ... R_{2j-2}^4}[f_{2j-1}(R_{2j-1},
 v_{2j-2},v_{2j-1})
\nonumber \\
&+&{R_{2j-1}^4\over R_{2j}^4}f_{2j}(R_{2j}, v_{2j}, v_{2j-1})]
+... ,\label{pot}
\end{eqnarray}
where $f$ is defined as
\be
f(R,u,v)= {(\nu+2)(R^{\nu} u-R^2 v)^2+(\nu-2)(u-R^{\nu+2} v)^2\over
1-R^{2\nu}}.
\ee

The effective potential in (14) is an iterated function with many
variables, however
it can be shown that a minimum exists in certain
conditions.
Defining that $r_{(2j-2,2j-1)}\equiv v_{2j-2}/v_{2j-1}$,
$r_{(2j,2j-1)}\equiv v_{2j}/v_{2j-1}$,
we obtain that
\begin{eqnarray}
\tilde{V}_{2j-1} & \equiv & V_{2j-1} + {R_{2j-1}^4\over R_{2j}^4}
({v_{2j+1}\over v_{2j-1}})^2 \tilde{V}_{2j+1} \cr\nonumber \\
&=&f_{2j-1}(R_{2j-1}, r_{(2j-2,2j-1)}, 1)+{R_{2j-1}^4\over R_{2j}^4}f_{2j}
(R_{2j}, r_{(2j,2j-1)}, 1) \cr\nonumber \\
&+&{R_{2j-1}^4\over R_{2j}^4}
({v_{2j+1}\over v_{2j-1}})^2 \tilde{V}_{2j+1} .
\end{eqnarray}

Substituting (15) into (16)
$\tilde{V}_{2j-1}$ can be rewritten as

\begin{eqnarray}
 &&\tilde{V}_{2j-1}=[\frac{(\nu+2)(R_{2j-1}^{\nu}
r_{(2j-2,2j-1)} - R_{2j-1}^2)^2+(\nu -2)(r_{(2j-2,2j-1)}
-R_{2j-1}^{\nu+2})^2}{1-R_{2j-1}^{2\nu}}]+\\
&& \frac{R_{2j-1}^4}{R_{2j}^4}
 [\frac{(\nu+2)(R_{2j}^2-R_{2j}^{\nu}r_{(2j,2j-1)})^2+(\nu-2)
(r_{(2j,2j-1)}-R_{2j}^{\nu+2})^2}{1-R_{2j}^{2\nu}}
+\tilde{V}_{2j+1}(\frac{v_{2j+1}}{v_{2j-1}})^2].
\end{eqnarray}

From $(17)$, one can see that for arbitrary positive
values of $\nu$, $\tilde{V}_{2j-1}$
grows as $R_{2j-1}\rightarrow 1$ or as $R_{2j}\rightarrow 1$
as long as $v_{2j-1}\not= v_{2j-2}$ and
$v_{2j-1} \not= v_{2j}$. These two limits correspond to the $(2j-1)$th
$"-"$ brane
approaching the $"+"$ brane at $y=y_{2j-2}$ and at $y=y_{2j}$,
respectively, and in these limits $\tilde{V}_{2j-1}$ is
singular, $\it i.e.$
$\tilde{V}_{2j-1}(R_{2j-1},R_{2j})\sim (1-R_{2j-1}^{2\nu})^{-1}>0$
as $R_{2j-1}\rightarrow 1$ and $\tilde{V}_{2j-1}(R_{2j-1},R_{2j})
\sim (1-R_{2j}^{2\nu})^{-1}>0
$ as $R_{2j}\rightarrow 1$. This implies that the $(2j-1)$th
$"-"$ brane experiences repulsive forces exerted on it by the
$"+"$ brane of its either side
and consequently the numbers of the branes can not be reduced.
We note that when $v_{2j-1}= v_{2j-2}$ and/or $v_{2j-1} = v_{2j}$,
 the leading singularity in $\tilde{V}_{2j-1}$ is removed and the
subleading terms in $\tilde{V}_{2j-1}$ is attractive.
In this case, therefore, the less brane crystals will be more stable.

From now on, we assume that
$v_{i}$ takes different numerical values in the different branes.
To obtain the values of $R_{2j-1}$ and $R_{2j}$, we minimize the effective
potential $\tilde{V}_{2j-1}$.
For $R_{2j}$ it satisfies the following equation:
\be
r_{(2j,2j-1)}^\pm (R_{2j}) =
{\frac{\nu\,{R_{2j}^{2 + \nu}}\,
            \left[ \left( 2 \pm {\sqrt{Q_{2j}}} \right) \,
          \left( {R_{2j}^{2\,\nu}} - 1 \right)    +
       \nu\,\left( 1 + {R_{2j}^{2\,\nu}} \right)  \right] }{2\,
     \left( {{\nu}^2}\,{R_{2j}^{2\,\nu}} +
       2\,{{\left( {R_{2j}^{2\,\nu}} - 1\right) }^2} +
       \nu\,\left({R_{2j}^{4\,\nu}} - 1 \right)  \right) }},
\label{Osoln}
\ee
where
\be
Q_{2j} = \nu^2 - 4 + 4 \tilde{V}_{2j+1}({v_{2j+1}\over v_{2j-1}})^2
\left[{(\nu+2)R_{2j}^{4\nu}+(\nu^2-4)R_{2j}^{2\nu}+2-\nu \over \nu^2
R_{2j}^{2\nu+4}}\right].
\ee
Since $R_{2j}\rightarrow 0$ or $1$,
$\tilde{V}_{2j-1} \rightarrow \infty$ which one can see from (17), this
extremum represents the minima of $\tilde{V}_{2j-1}$ in the $R_{2j}$-direction.
To see whether simultaneous minima in the $R_{2j-1}$-direction
exist, we extremize $\tilde{V}_{2j-1}$ with respect to $R_{2j-1}$ and get
\be
r_{(2j-2,2j-1)}^\pm (R_{2j-1}) = \frac{R_{2j-1}^{2 - \nu}}{2\, \nu} \,
   \left[ 2\, \left( 1 - {R_{2j-1}^{2\,\nu}} \right)
     + {\nu}\, \left( 1 + {R_{2j-1}^{2\,\nu}} \right)
\pm
    R_{2j-1}^{2 + \nu} ( 1 - R_{2j-1}^{2 \nu}) \sqrt{Q_{2j-1}}
            \right],
   \,
\ee
where
\be
Q_{2j-1} =\nu^2 - 4 - 4\left[{f_{2j}(R_{2j}, r_{(2j,2j-1)}, 1)\over R_{2j}^4}
+{\tilde{V}_{2j+1} v_{2j+1}^2 \over R_{2j}^4 v_{2j-1}^2} \right].
\ee
Note that for $r_{(2j-2,2j-1)} = r_{(2j-2,2j-1)}^+$,
$\partial^2 \tilde{V}_{2j-1} / \partial R_{2j-1}^2 < 0 $
, which
corresponds to a sequence of saddle points.
While for $r_{(2j-2,2j-1)} = r_{(2j-2,2j-1)}^-$,
$\partial^2 \tilde{V}_{2j-1}/ \partial R_{2j-1}^2 > 0 $
and this extremum represents the minima of $\tilde{V}_{2j-1}$ in
the $R_{2j-1}$-direction.
Thus in the parameter space where $r_{(2j,2j-1)}^\pm$ and
$r_{(2j-2,2j-1)}^-$ co-exist,
the absolute minima of $\tilde{V}_{2j-1}$
exists.
Therefore, the minima of $V(R_i,r_i)$
exists, i.e. Crystal Universe can be stabilized.

Having shown the possibility of stabilizing the Crystal Universe
models, we discuss and analyze the configuration of the
brane crystals
when they are stabilized. Following GW\cite{GW}, we will also limit
ourselve to
the regime where $\epsilon$ is small,
$\epsilon \equiv \nu - 2 \approx \frac{m^2}{4k^2}\ll 1$.
 For a stabilized Crystal Universe model, from $(21)$ we have
\be
{f_{2j}(R_{2j},
r_{(2j,2j-1)}, 1)\over R_{2j}^4}
+{\tilde{V}_{2j+1} \over R_{2j}^4}({v_{2j+1}\over v_{2j-1}})^2 \leq
\epsilon.
\ee
Note that the two terms on the left-handed side of eq.(22) are positive,
we have seperately
\be
{f_{2j}(R_{2j},
r_{(2j,2j-1)}, 1)\over R_{2j}^4}\leq \epsilon,
\label{Rsoln}
\ee

\be
{\tilde{V}_{2j+1}\over R_{2j}^4}({v_{2j+1}\over v_{2j-1}})^2 \leq \epsilon.
\ee
In Fig.2 we plot the allowed region of $R_{2j}$ based on (23), from which
one can see
that for $\epsilon = 0.01$
$R_{2j}$ varies from
$0.85$ to 1, which
corresponds to the $(2j-1)$th $"-"$ brane very close to the $2j$th $"+"$
brane of its right side.
With $R_{2j}$ in this range  we have $r_{(2j,2j-1)}\approx 1$,
i.e. $v_{2j}\approx v_{2j-1}$.

From eq. $(24)$ we get
\be
({v_{2j+1}\over v_{2j-1}})^2{V_{2j+1}\over R_{2j}^4}
+{R_{2j+1}^4\over R_{2j}^4 R_{2j+2}^4}({v_{2j+1}\over v_{2j-1}})^2
({v_{2j+3}\over v_{2j+1}})^2 \tilde{V}_{2j+3} \leq \epsilon.
\ee

Thus
\be
({v_{2j+1}\over v_{2j-1}})^2{V_{2j+1}\over R_{2j}^4}\leq \epsilon,
\ee

\be
{R_{2j+1}^4\over R_{2j}^4 R_{2j+2}^4}({v_{2j+1}\over v_{2j-1}})^2
({v_{2j+3}\over v_{2j+1}})^2 \tilde{V}_{2j+3}  \leq \epsilon.
\ee

Given that $v_{2j-1}\approx v_{2j}$, we have from (26)
\be
{f_{2j+1}(R_{2j+1},r_{(2j,2j+1)},1)\over r_{(2j,2j+1)}^2 }\leq \epsilon R_{2j}^4 ,
\ee
\be
{1\over r_{(2j,2j+1)}^2}{R_{2j+1}^4\over R_{2j}^4 R_{2j+2}^4}
f_{2j+2}(R_{2j+2},r_{(2j+2,2j+1)},1)\leq \epsilon .
\ee
In Fig.3 we plot the allowed range of $R_{2j+1}$ from which we see that
with $R_{2j}$ in the range of $0.85\sim 1$, $R_{2j+1}$ varies from
$0$ to $0.8 \sim 1$





Now we consider an additonal constraint on $R_{2j+1}$ from (27).
Combining
Eqs.$(26)$, $(27)$ and $(29)$ we obtain
\be
R_{2j+1}^4/r_{(2j,2j+1)}^2 \leq R_{2j+2}^4,
\ee
which we plot in Fig.4 for the allowed range of $R_{2j+1}$.

From Figs. 3 and 4 one can see that a Crystal Universe can be stabilized
for a large parameter space of
$R_{2j+1}$, however
$R_{2j}$ is required to be very close to $1$.


Similarly we can discuss the parameter space for solution (ii) and our
results show to have a stabilized Crystal Universe
$R_{2j+1}$ approachs $1$, and correspondingly $r_{(2j+1,2j)}\approx 1$.
However we should point out that
the absolute value of
$R_1$ can not be fixed and it depends mostly on the $r_{(1,0)}$.


In summary,
in this paper by explicit calculation we show that
Crystal Universe can be stabilized by introducing a bulk
scalar field to brane system. Our results differ from the assumptions
taken in the literature for the discussion of the Crystal Universe. For
exmple, in Refs.\cite{KM1},\cite{K}.  they have assumed that the branes are equidistant.

We should point out that Choudhury et al. \cite{CJ}
studied the stability of
the $"+-+"$ brane configuration and find that the $"-"$
brane chooses to stay close to the visible $"+"$ brane. Taking $n=2$, we
recover the results of Ref.\cite{CJ}. For $n=1$ we agree with GW's
results. So our results apply for general Crystal Universe models.

\textbf{Acknowledgments}

We thank Prof. Y.L. Wu for stimulating discussion. We also thank
J.Y. Fan, D.H. Hang, W.B. Lin, F. Liu for discussions. This work is
supported in part by National Natural Science Foundation of
China under
Grant Nos. 10047004 and 19835040, and also supported by Ministry of Science and Technology of Modern Grant No. NKBRSF G19990754.

\begin{figure}
\begin{center}
\setlength{\unitlength}{0.1in}
\begin{picture}(40,8)

\put(0,5){\line(1,0){8}}
\put(10,5){\line(1,0){10}}
\put(22,5){\line(1,0){6}}
\put(30,5){\line(1,0){2}}
\put(32.5,5){\line(1,0){0.5}}
\put(33.5,5){\line(1,0){0.5}}
\put(34.5,5){\line(1,0){0.5}}
\put(35.5,5){\line(1,0){0.5}}
\put(36.5,5){\line(1,0){0.5}}
\put(37.5,5){\line(1,0){0.5}}
\put(38.5,5){\line(1,0){1.5}}

\put(-1,5){\circle{2}}
\put(9,5){\circle * {2}}
\put(21,5){\circle{2}}
\put(29,5){\circle * {2}}
\put(41,5){\circle{2}}

\put(-1.5,2){\makebox(1,1)[c]{$+$}}
\put(8.5,2){\makebox(1,1)[c]{$-$}}
\put(20.5,2){\makebox(1,1)[c]{$+$}}
\put(28.5,2){\makebox(1,1)[c]{$-$}}
\put(40.5,2){\makebox(1,1)[c]{$+$}}

\end{picture}
\end{center}
\caption{Crystal Universe is made up $n+1$ array of $"+"$ and
$"-"$ branes
with lattice
spacing $(y_1-y_0)$, $(y_2-y_1)$ ... $(y_n-y_{n-1})$
and bulk curvature $k$. The fifth dimension $y$ has orbifold
geometry $S^1/Z_2$, $y_0=0$ and $y_n$ are orbifold fixed points.}

\end{figure}
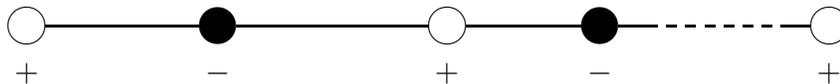

\begin{figure}[ht]
\begin{center}
\mbox{\epsfig{file=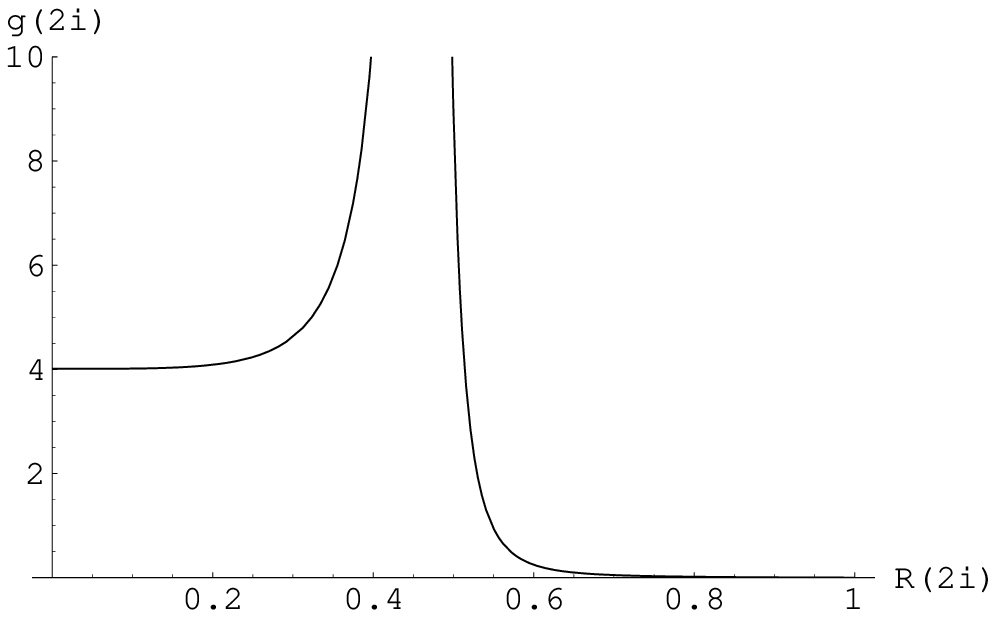,width=6cm}\epsfig{file=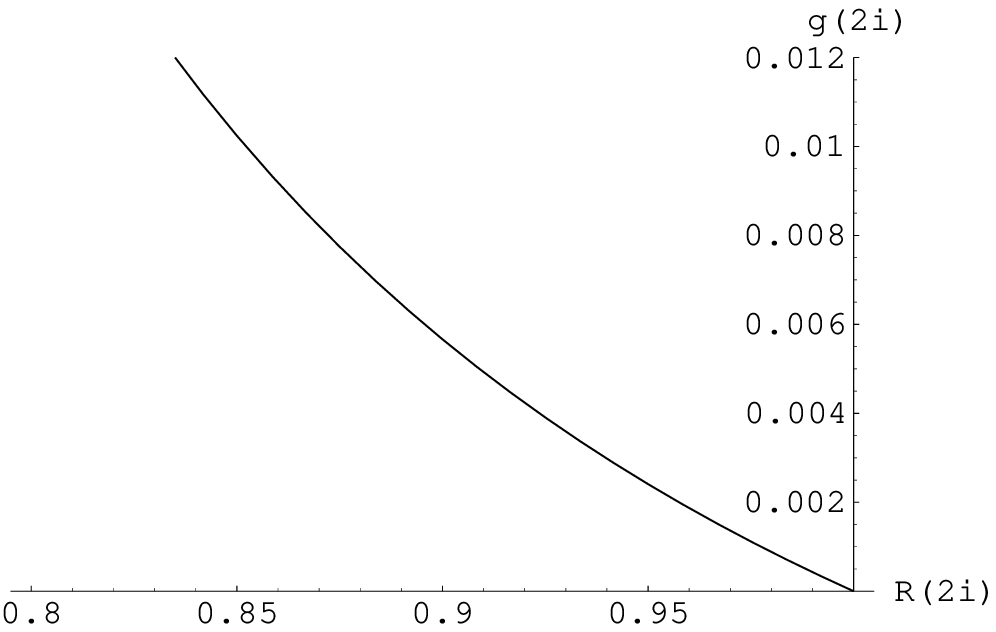,width=6cm}}
\caption {The allowed parameter space of $R_{2j}$. The y-axis is
  $g_{2j}\equiv{f_{2j}\over R_{2j}^4}$ {\it i.e.} the left-handed side of
eq.(23); the x-axis is $R_{2j}$.
The figure on the right-handed side is an amplification of the left in the
range of $R_{2j} \simeq 0.85\sim 1$.
}
\end{center}
\end{figure}

\begin{figure}

\begin{center}

\mbox{\epsfig{file=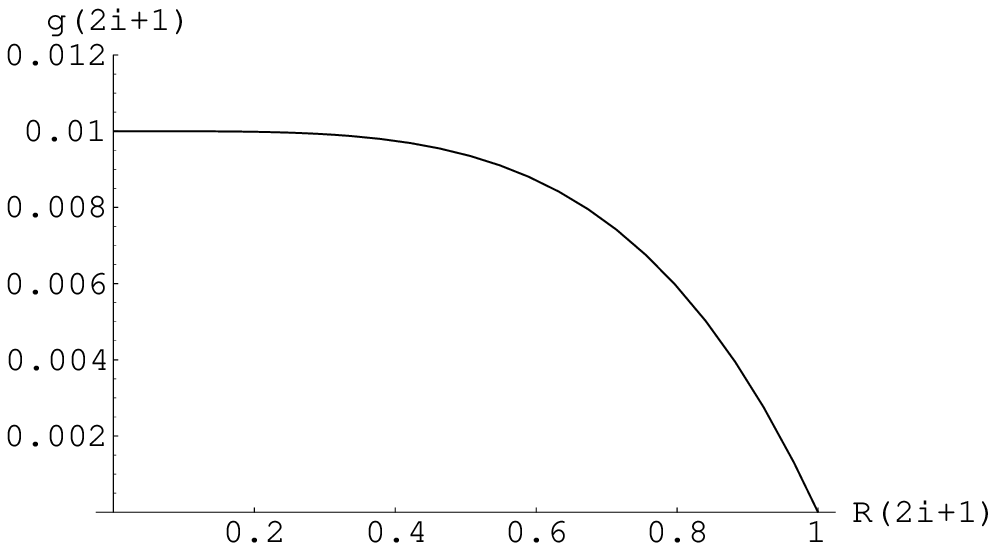,width=6cm}}
\caption { The allowed parameter space of $R_{2j}$. The y-axis is
  $g_{2j}\equiv{f_{2j}\over r_{(2j,2j+1)}^2}$ {\it i.e.} the left-handed side of
eq.(28); the x-axis is $R_{2j+1}$.}

\end{center}
\end{figure}

\begin{figure}
\begin{center}
\mbox{\epsfig{file=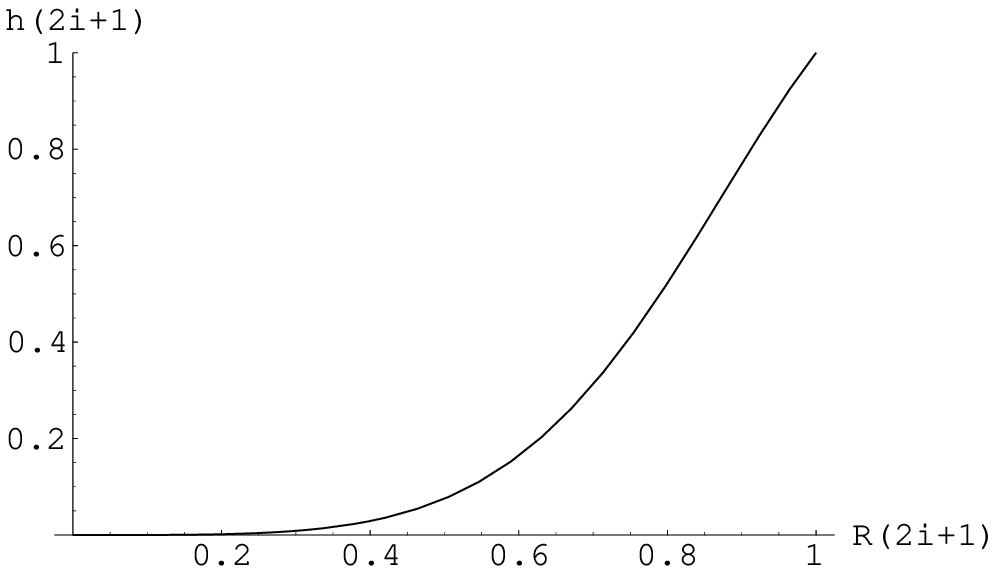,width=6cm}}
\caption { The allowed parameter space of $R_{2j}$. The y-axis is
  $h_{2j}\equiv{R_{2j+1}\over r_{(2j,2j+1)}^4}$ {\it i.e.} the left-handed side of
eq.(30); the x-axis is $R_{2j+1}$.
}
\end{center}

\end{figure}

\begin{figure}
\begin{center}
\setlength{\unitlength}{0.1in}
\begin{picture}(40,20)

\put(-7,5){\line(1,0){5}}
\put(0,5){\line(1,0){1}}
\put(3,5){\line(1,0){4}}
\put(9,5){\line(1,0){1}}
\put(12,5){\line(1,0){6}}
\put(20,5){\line(1,0){1}}
\put(23,5){\line(1,0){2}}

\put(25.5,5){\line(1,0){0.5}}
\put(26.5,5){\line(1,0){0.5}}
\put(27.5,5){\line(1,0){0.5}}
\put(28.5,5){\line(1,0){0.5}}
\put(29.5,5){\line(1,0){0.5}}
\put(30.5,5){\line(1,0){0.5}}

\put(31.5,5){\line(1,0){1.5}}
\put(35,5){\line(1,0){1}}
\put(38,5){\line(1,0){5}}

\put(-8,5){\circle {2}}
\put(-1,5){\circle*{2}}
\put(2,5){\circle  {2}}
\put(8,5){\circle*{2}}
\put(11,5){\circle  {2}}
\put(19,5){\circle*{2}}
\put(22,5){\circle {2}}
\put(34,5){\circle*  {2}}
\put(37,5){\circle{2}}
\put(44,5){\circle* {2}}

\put(-8.5,2){\makebox(1,1)[c]{$+$}}
\put(-1.5,2){\makebox(1,1)[c]{$-$}}
\put(1.5,2){\makebox(1,1)[c]{$+$}}
\put(7.5,2){\makebox(1,1)[c]{$-$}}
\put(10.5,2){\makebox(1,1)[c]{$+$}}
\put(18.5,2){\makebox(1,1)[c]{$-$}}
\put(21.5,2){\makebox(1,1)[c]{$+$}}

\put(33.5,2){\makebox(1,1)[c]{$-$}}
\put(36.5,2){\makebox(1,1)[c]{$+$}}
\put(43.5,2){\makebox(1,1)[c]{$-$}}

\put(-9.5,8){\makebox(3,1)[c]{$y_0 =0$}}
\put(43.5,8){\makebox(1,1)[c]{$y_n$}}

\put(-7,13){\line(1,0){5}}
\put(0,13){\line(1,0){1}}
\put(3,13){\line(1,0){4}}
\put(9,13){\line(1,0){1}}
\put(12,13){\line(1,0){6}}
\put(20,13){\line(1,0){1}}
\put(23,13){\line(1,0){2}}

\put(25.5,13){\line(1,0){0.5}}
\put(26.5,13){\line(1,0){0.5}}
\put(27.5,13){\line(1,0){0.5}}
\put(28.5,13){\line(1,0){0.5}}
\put(29.5,13){\line(1,0){0.5}}
\put(30.5,13){\line(1,0){0.5}}

\put(31.5,13){\line(1,0){1.5}}
\put(35,13){\line(1,0){1}}

\put(-8,13){\circle {2}}
\put(-1,13){\circle*{2}}
\put(2,13){\circle  {2}}
\put(8,13){\circle*{2}}
\put(11,13){\circle  {2}}
\put(19,13){\circle*{2}}
\put(22,13){\circle {2}}
\put(34,13){\circle*  {2}}
\put(37,13){\circle{2}}

\put(-9.5,16){\makebox(3,1)[c]{$y_0 =0$}}
\put(36.5,16){\makebox(1,1)[c]{$y_n$}}

\end{picture}
\end{center}
\caption{Illustration of a Crystal
Universe model with  $"+"$ brane at
orbifold fixed point $y_0=0$. The up figure ends with $"+"$
brane and the down with $"-"$ brane.}
\end{figure}

\begin{figure}
\begin{center}
\setlength{\unitlength}{0.1in}
\begin{picture}(40,20)

\put(-7,5){\line(1,0){5}}
\put(0,5){\line(1,0){7}}
\put(9,5){\line(1,0){1}}
\put(12,5){\line(1,0){6}}
\put(20,5){\line(1,0){1}}
\put(23,5){\line(1,0){2}}

\put(25.5,5){\line(1,0){0.5}}
\put(26.5,5){\line(1,0){0.5}}
\put(27.5,5){\line(1,0){0.5}}
\put(28.5,5){\line(1,0){0.5}}
\put(29.5,5){\line(1,0){0.5}}
\put(30.5,5){\line(1,0){0.5}}

\put(31.5,5){\line(1,0){1.5}}
\put(35,5){\line(1,0){1}}
\put(38,5){\line(1,0){5}}

\put(-8,5){\circle* {2}}
\put(-1,5){\circle {2}}
\put(8,5){\circle*{2}}
\put(11,5){\circle  {2}}
\put(19,5){\circle*{2}}
\put(22,5){\circle {2}}
\put(34,5){\circle*  {2}}
\put(37,5){\circle{2}}
\put(44,5){\circle* {2}}

\put(-8.5,2){\makebox(1,1)[c]{$-$}}
\put(-1.5,2){\makebox(1,1)[c]{$+$}}
\put(7.5,2){\makebox(1,1)[c]{$-$}}
\put(10.5,2){\makebox(1,1)[c]{$+$}}
\put(18.5,2){\makebox(1,1)[c]{$-$}}
\put(21.5,2){\makebox(1,1)[c]{$+$}}

\put(33.5,2){\makebox(1,1)[c]{$-$}}
\put(36.5,2){\makebox(1,1)[c]{$+$}}
\put(43.5,2){\makebox(1,1)[c]{$-$}}

\put(-9.5,8){\makebox(3,1)[c]{$y_0 =0$}}
\put(43.5,8){\makebox(1,1)[c]{$y_n$}}

\put(-7,13){\line(1,0){5}}
\put(0,13){\line(1,0){7}}
\put(9,13){\line(1,0){1}}
\put(12,13){\line(1,0){6}}
\put(20,13){\line(1,0){1}}
\put(23,13){\line(1,0){2}}

\put(25.5,13){\line(1,0){0.5}}
\put(26.5,13){\line(1,0){0.5}}
\put(27.5,13){\line(1,0){0.5}}
\put(28.5,13){\line(1,0){0.5}}
\put(29.5,13){\line(1,0){0.5}}
\put(30.5,13){\line(1,0){0.5}}

\put(31.5,13){\line(1,0){1.5}}
\put(35,13){\line(1,0){1}}

\put(-8,13){\circle *{2}}
\put(-1,13){\circle{2}}
\put(8,13){\circle*{2}}
\put(11,13){\circle  {2}}
\put(19,13){\circle*{2}}
\put(22,13){\circle {2}}
\put(34,13){\circle*  {2}}
\put(37,13){\circle{2}}

\put(-9.5,16){\makebox(3,1)[c]{$y_0 =0$}}
\put(36.5,16){\makebox(1,1)[c]{$y_n$}}

\end{picture}
\end{center}
\caption{Illustration of a Crystal
Universe model with  $"-"$ brane at
orbifold fixed point $y_0=0$. The up figure ends with $"+"$
brane and the down with $"-"$ brane.}
\end{figure}

\end{document}